\documentclass[useAMS,usenatbib,onecolumn,usegraphicx]{mn2e}
\usepackage{amssymb}
\usepackage{color}

\title[Non-exponential  growth in   Keplerian discs]
{Non-exponential hydrodynamical growth in density-stratified thin
Keplerian discs}
\author[Yu. M. Shtemler, M. Mond, G. R$\ddot{u}$diger,
 O. Regev, and  O. M. Umurhan]
{Yu. M. Shtemler$^{1}$\thanks{E-mail: shtemler@bgu.ac.il;
mond@bgu.ac.il; gruediger@aip.de; regev@astro.columbia.edu;
umurhan@maths.qmul.ac.uk},
 M. Mond$^{1}$, G.
 R$\ddot{u}$diger$^{2}$,  O. Regev$^{3}$,  O. M. Umurhan$^{4}$\\
$^{1}$Department of Mechanical Engineering,  Ben-Gurion University
of the Negev,  P.O. Box 653, Beer-Sheva 84105,
Israel\\
$^{2}$Astrophysikalisches Institut Potsdam, An der Sternwarte 16,
D-14482 Potsdam, Germany\\
$^{3}$Department of Physics, Technion – Israel Institute of
Technology, 32000 Haifa, Israel \\and Department
of Astronomy, Columbia University, New York, New York 10027, USA\\
$^{4}$ Astronomy Unit, School of Mathematical Sciences, Queen Mary
University of London,\\ Mile End Road, London E1 4NS, United
Kingdom\\ and Department of Astronomy, City College San Francisco,
San Francisco,
California 94112, USA\\
 }
\begin{document}

\date{Accepted ---. Received ----; in original form ----}

\pagerange{\pageref{firstpage}--\pageref{lastpage}} \pubyear{}

\maketitle

\label{firstpage}

\begin{abstract}
The short time evolution of three dimensional small perturbations
is studied. Exhibiting spectral asymptotic stability, thin discs
are nonetheless shown to host intensive hydrodynamical activity in
the shape of non modal growth of initial small perturbations. Two
mechanisms that lead to such behavior are identified and studied,
namely, non-resonant excitation of vertically confined sound waves
by stable planar inertia-coriolis modes that results in linear
growth with time, as well as resonant coupling of those two modes
that leads to a quadratic growth of the initial perturbations. It
is further speculated that the non modal growth can give rise to
secondary strato-rotational instabilities and thus lead to a new
route to turbulence generation in thin discs.
\end{abstract}


\section{Introduction}
Reverting to Rayleigh's stability criterion Keplerian accretion
discs have been deemed centrifugally stable. This has lead to
focusing the attention on the magneto-rotational instability (MRI)
which is excited due to the freezing of magnetic field lines into
fluid elements \cite{Balbus and Hawley 1991}. Indeed, the MRI has
been accepted since as a main turbulence generator and has
consequently been widely considered to hold the key to the
solution of the angular transfer problem in accretion discs.
However, the MRI as a driver of turbulence and angular momentum
transport has recently been shown to suffer from non-trivial
difficulties in the sheer use of the shearing box in the
simulations [\cite{Regev and Umurhan 2008}, \cite{Bodo et al.
2008}] as well as in doubts regarding the numerical convergence
and resolution [\cite{Lesur and Longaretti 2007}, \cite{Fromang
and Papaloizou 2007}, \cite{Fromang et al. 2007}, \cite{Pessah et
al. 2007}]. In addition, it has recently been shown that if the
thin disk geometry is taken into account both the growth rates as
well as the number of unstable MRI modes are greatly reduced and
are decreasing functions of the disk thickness [\cite{Coppi and
Keyes 2003}, \cite{Liverts and Mond 2009}]. In parallel, various
works have indicated over the last decade that pure hydrodynamical
processes are not as irrelevant as possible sources for enhanced
transport coefficients in thin accretion discs as has been hastily
assumed before. In particular, two topics have caught the
attention of researchers. In the first one the destabilizing
effect of a stable stratification on a centrifugally stable
rotating fluid has been examined. The resulting instability,
termed the strato-rotational instability (SRI), has been
investigated for Taylor-Couette flows [\cite{Boubnov et al. 1995},
\cite{Yavneh et al. 2001}, \cite{Shalybkov  and Rudiger 2005}] or
for Taylor-Couette-like flows within the shearing sheet
approximation that models rotating discs [\cite{Dubrulle et al.
2005}, \cite{Umurhan 2006}, \cite{Tevzadze et al. 2008}]. Common
to those works is: 1. Rigid-like boundary conditions at two radial
locations, 2. Radial pressure distribution plays a crucial role in
the onset of the SRI under those conditions, 3. Ignoring the
vertical boundaries of the fluid domain, and 4. Incompressibility
within the Boussinesq approximation. We will return later on to
these points (actually 1 and 2 are two sides of the same
phenomenon) but let us first move on to the second route that  may
lead to pure hydrodynamical turbulence. The latter is motivated by
successful attempts by fluid dynamics to explain the transition to
turbulence in spectrally stable flows [\cite{Boberg and Brosa
1988}, \cite{Grossmann 2000}, \cite{Schmid and Henningson 2000}].
Thus, rendering the linear operator non normal, the flow shear may
give rise to transiently growing perturbations [{\cite{Gustavsson
1991}, \cite{Butler and Farrel 1992}, \cite{Reddy and Henningson
1993}, \cite{Trefethen et al. 1993}, \cite{Criminale et al. 1997},
and the recent review by \cite{Schmid 2007}] whose energy is
eventually redistributed among different length-scales due to
nonlinear effects. Indeed, recent calculations in real thin disc
geometry have demonstrated the efficiency of such processes to
significantly amplify initially small perturbations, and thus to
generate intensive hydrodynamical activity in the otherwise
centrifugally stable Keplerian discs [\cite{Umurhan et al. 2006} ,
\cite{Rebusco et al. 2009}]. Such non modal growth of
perturbations has also been obtained for ballooning modes in
magnetized rotating plasmas in tokamaks [\cite{Chun  and Hameiri
1990}]. The two routes described above are shown in the current
work to diverge from a single unified comprehensive stability
analysis in real thin disc geometry. It is thus shown first that
if the proper physical conditions as well as the true thin disc
geometry are taken into account, no spectrally unstable
strato-rotation modes exist. Indeed, the lack of radial rigid
boundaries, and the negligible role of the radial pressure
gradients in establishing the steady-state rotation stand in stark
distinction from the conditions that give rise to the SRI in
Couette flows. Furthermore, the small thickness of the disc
combined with its steady-state supersonic rotation results in
vertical sound crossing time that is of the order of a rotation
period. Consequently, compressibility effects cannot be ignored so
that vertically propagating sound waves play a major role in the
dynamical response of thin discs to small perturbations.

Thus,  (temporary) disappearing from the thin discs global scene
 (they will resurface later on), the SRI leaves
the stage entirely to algebraic instabilities as a primary source
of hydrodynamical activity. As will be shown there are two
mechanisms that give rise to such non-exponential dynamical
response. One is the non-resonant coupling between the two modes
of wave propagation in the system (the inertia-coriolis waves and
the sound waves). Such coupling is a direct result of the rotation
shear that renders the linear operator non normal. Due to such
interaction, sound waves are non-resonantly driven by the
inertia-coriolis modes and their amplitude grows linearly in time.
The second source of non-exponential perturbation growth is the
resonant interaction between the above mentioned modes. In such
scenario, sound waves are resonantly driven by the
inertia-coriolis modes and due to the shear their amplitude grows
quadratically in time.

Finally, the two routes converge again by the tendency of the
algebraically growing perturbations to develop sharp radial
pressure gradients due to the rotation shear. As such gradients
are essential ingredients in the onset of the SRI, it is
conjectured that the latter may be excited as a secondary
instability on top of the non modal growth.

The paper is organized as follows. The dimensionless governing
equations and their approximation to leading order in small aspect
ratio of the steady-state disc are presented in the next Section.
Section 3 describes the principal hydrodynamic modes of waves
which can propagate in Keplerian discs. Non-resonantly as well as
resonantly driven sound waves by inertial waves are considered in
Sections 4 and 5, respectively. Conclusions are presented in
section 6. Appendix A contains the principal boundary-value
problems and methods of their solutions.

\section{The governing equations }
The dynamical response of radially and axially stratified thin
rotating fluid discs to small perturbations is considered. For
that purpose it is convenient to transform all the physical
variables to non-dimensional quantities by using the following
characteristic values [\cite{Shtemler et al. 2007}, (2009)]:
%
\begin{equation}
V_*=\Omega_* r_* ,\,\,t_*=\frac{1}{\Omega_*},\,\,
\Phi_*={V_*}^2,\,\, m_*= m_i,\,\,
n_*=n_i,\,\,\,
 P_*=K(m_* n_*)^\gamma.\,\,
\,\,\,\,\,\,\,\,\,\,\,\,\,\,\,\,\,\,\,\,\,\,\,\,\,\,\,\,\,\,\,\,\,\,\,\,\,\,
\label{1}
\end{equation}
Here $\Omega_*=(GM_c/r_*^3)^{1/2}$  is the Keplerian angular
velocity of the fluid at the characteristic radius $r_*$ that
belongs to the Keplerian portion of the disc ;
$G$  is the gravitational constant; $M_c$ is the total mass of the
central object; $\Phi_*$  is the characteristic value of the
gravitational potential; $m_*$  and $n_*$   are the characteristic
mass and number density;
%
 %
$m_i$  and $n_i$ are the ion mass and number density;
$K$ is the dimensional constant in the
polytropic law $P=Kn^\gamma$, $\gamma$ is the polytrophic
coefficient, ($\gamma=5/3$ in the adiabatic case).
%
The resulting dimensionless dynamical equations, expressed in
standard cylindrical coordinates $\{r,\phi,z\}$  are:
\begin{equation}
\frac{\partial V_r}{\partial t}+ V_r \frac{\partial V_r}{\partial
r} +\frac{ V_\phi}{ r}\frac{\partial V_r}{\partial \phi } + V_z
\frac{\partial V_r}{\partial z}-\frac{ V_\phi^2}{ r} =
-\frac{1}{M_S^2}\frac{1}{n}\frac{\partial P}{\partial
r}-\frac{\partial \Phi(r,z)}{\partial r},
\label{2}
\end{equation}
\begin{equation}
\frac{\partial V_\phi }{\partial t}+ V_r \frac{\partial V_\phi
}{\partial r} +\frac{ V_\phi}{ r}\frac{\partial V_\phi }{\partial
\phi } + V_z \frac{\partial V_\phi }{\partial z}+\frac{V_r
V_\phi}{ r} =- \frac{1}{M_S^2}\frac{1}{n r}\frac{\partial
P}{\partial \phi },
\label{3}
\end{equation}
\begin{equation}
\frac{\partial V_z}{\partial t}+ V_r \frac{\partial V_z}{\partial
r} +\frac{ V_\phi}{ r}\frac{\partial V_z}{\partial \phi } + V_z
\frac{\partial V_z}{\partial z } =
-\frac{1}{M_S^2}\frac{1}{n}\frac{\partial P}{\partial
z}-\frac{\partial \Phi(r,z)}{\partial z},
\label{4}
\end{equation}
\begin{equation}
\frac{\partial n}{\partial t}+ \frac{1}{r}\frac{\partial r n V_r}{\partial r}
+\frac{1}{r}\frac{\partial n V_\phi}{ \partial \phi } +
\frac{\partial nV_z}{\partial z }=0,
\,\,\,\,\,\,\,\,\,\,\,\,\,\,\,
\label{5}
\end{equation}
\begin{equation}
P =n^{\gamma}.
\label{6}
\end{equation}
Here the Mach number is defined as $M_S=V_*/C_{S*}$, the
characteristic sound velocity is $ C_{S*}=\sqrt{P_*/(m_*n_*}$, $t$
is time, $\{V_r, V_\phi, V_z\}$   is the plasma velocity,  $P$ and
$n$ are the pressure and density current, and $\Phi$ is the
gravitational potential due to the central object. Note that a
preferred direction is tacitly defined here, namely, the positive
direction of the $z$ axis is chosen according to positive
Keplerian rotation.

A common property of thin Keplerian discs is their highly
compressible motion with large Mach numbers [ \cite{Frank et al.
2002}]:
\begin{equation}
\frac{1}{M_S}=\epsilon=\frac{H_*}{r_*}
\ll 1,
\label{8}
\end{equation}
where the characteristic effective semi-thickness $H_*$  of the
equilibrium disc is such that the disc aspect ratio $\epsilon$
equals the inverse Mach number (for the magnetic-field-free
systems under consideration, the value$H_*$  should be fixed and
known). The smallness of  $\epsilon$ means that dimensionless
axial coordinate is also small, i.e. $z/r_*\sim \epsilon$  ($\mid
z\mid \leq H_*$), and consequently it is convenient to introduce
the following rescaled values in order to further apply the
asymptotic expansions in $\epsilon$ [see \cite{Regev 1983},
\cite{Ogilvie 1997}, \cite{Kluzniak and Kita 2000}, \cite{Regev
and Gitelman 2002}, \cite{Shtemler et al. 2007}, (2009)]:
\begin{equation}
\zeta=\frac{z}{\epsilon}\sim\epsilon^0,\,\,\,H_{\pm 1} =\pm
\frac{H_*}{\epsilon}\sim\epsilon^0,\,\,\,V_{z1}=\frac{V_z}{\epsilon}\sim\epsilon^0,
\label{9}
\end{equation}
where the vertical velocity is scaled along with the disc height
in $\epsilon$  as follows from the principle of the least
degeneracy of the problem for small  $\epsilon$.

The set of Eqs. (\ref{2})-(\ref{6})  is complemented with the
conventional dynamic and kinematical boundary conditions on the
interfaces, according to which the disc edges $z=H_{\pm}(t,r,
\phi, z)$ are determined self-consistently. However, to simplify
the problem, it is assumed that the disc
 edges are given by  $z=\pm H_*\equiv \pm const$, where $H_*$ the characteristic
effective semi-thickness of the equilibrium disc. Consequently,
the kinematic boundary conditions are not needed,
while the dynamic  conditions $P=n^\gamma=0$ at the interfaces are
modelled by:
\begin{equation} n =0\,\,\,\mbox{for
}\,\,\,\zeta =\pm 1.
\label{7}
\end{equation}
Using of the model conditions (\ref{7}) instead of the true
 boundary conditions at the interfaces indeed leads to the loss
of the corresponding interface equations. However, this does not
lead to a change in the essential nature of the dynamical behavior
of the disc, and the kinematic boundary conditions provide
separate equations for calculating the perturbed of the top and
bottom interfaces of the disc.

The gravitational potential in  Eqs. (\ref{2}) and (\ref{4})   may
be expanded now in terms of the stretched axial variable to yield:
 \begin{equation}
\Phi(r,\zeta)=\Phi^{(0)}(r)+\epsilon^2\Phi^{(2)}(r,\zeta)+O(\epsilon^4),\,\,\,
\Phi^{(0)}(r)=-\frac{1}{r},\,\,\, \Phi^{(2)}(r,\zeta)
=\frac{\zeta^2}{2r^3},\,\,\, r>1>>\epsilon.
\label{10}
\end{equation}
Representing the solution of  Eqs. (\ref{2})-(\ref{6})   as a sum
of the equilibrium state and its perturbations
\begin{equation}
P=\bar{P}+P',\,\,\,n=\bar{n}+n',\,\,\,
V_r=V_r',\,\,\,V_\phi=\bar{V}_\phi+V_\phi',\,\,\,V_{z1}=V_z',\,\,\,
\label{11}
\end{equation}
the steady-state solution is given by
\begin{equation}
\bar{P}(r,\zeta)=\bar{n}^\gamma,\,\,\,
\bar{n}(r,\zeta)=\big(\frac{\gamma-1}{2\gamma}
\big)^{\frac{1}{\gamma-1}}\big(\frac{1-\zeta^2}{r^3}\big)^{\frac{1}{\gamma-1}}
,\,\,\, \bar{V}_\phi(r)=r\Omega(r),\,\,\, \Omega(r)=r^{-3/2}.
 \label{12}
\end{equation}
 Substituting  Eqs. (\ref{8})-(\ref{12}) into
(\ref{2})-(\ref{7}), linearizing about the steady-state solution,
and  employing the relation $P'=\bar{c}_S^2n'$, yield the
following linear set of equations for the perturbations to leading
order in $\epsilon$:
\begin{equation}
\frac{\partial V_r'}{\partial t}+\Omega(r)\frac{\partial
V_r'}{\partial \phi } -2\Omega  V_\phi' =0,
\label{13}
\end{equation}
\begin{equation}
\frac{\partial V_\phi' }{\partial t}
+\Omega(r)\frac{\partial V_\phi'}{\partial \phi }
+\frac{1}{2}\frac{\chi^2(r)}{\Omega(r)} V_r'
 =0,
\label{14}
\end{equation}
\begin{equation}
\frac{\partial \bar{n} V_z'}{\partial t}+ \Omega(r)\frac{\partial
\bar{n} V_z'}{\partial \phi }-\frac{2-\gamma}{\gamma-1}
\Omega^2(r)\frac{d \bar{c}_S^2}{d \zeta}n'+\Omega^2(r)\bar{c}_S^2\frac{\partial n'}{\partial
\zeta}=0,
\label{15}
\end{equation}
\begin{equation}
\frac{\partial n'}{\partial t}+ \Omega(r)\frac{\partial n' }{
\partial \phi } + \frac{\partial \bar{n}V_z'}{\partial z }=-\frac{1}{r}\frac{\partial r\bar{n} V_r'}{\partial r }
-\frac{1}{r}\frac{\partial \bar{n} V_\phi'}{\partial \phi },
\,\,\,\,\,\,\,\,\,\,\,\,\,\,\,
\label{16}
\end{equation}
where
 \begin{equation}
 \chi^2=2\frac{\Omega}{r}\frac{d(r^2\Omega)}{dr}, \,\,\,\,\,\,
 \bar{c}_S^2(\zeta)\equiv\frac{\bar{C}_S^2(r,\zeta)}{\Omega^2(r)}=\frac{\gamma-1}{2}(1-\zeta^2),
 \label{17}
\end{equation}
$\bar{C}_S=(\partial \bar{P}/\partial \bar{n})^{1/2}$ is the
dimensionless equilibrium sound velocity; $\chi$ is the epicyclic
frequency. For Keplerian discs the epicyclical frequency is equals
to Keplerian frequency $ \chi=\Omega(r)$.

Following Eq. (7), Eqs. (13)-(16) are supplemented with the
following boundary conditions:
 \begin{equation}
n' = 0  \,\,\,\mbox{for}\,\,\, \zeta=\pm 1.
 \label{18}
\end{equation}

\section{The Modes of Wave Propagation}
\subsection{Planar inertia-Coriolis waves}
We start the solution of the linearized system by noticing that
Eqs. (\ref{13}) and (\ref{14}) for $V_r'$  and $V_\phi'$   are
decoupled from the rest of the system due to the thin disc
approximation. Furthermore, the sub-set  (\ref{13})-(\ref{14})
represents a set of ordinary differential equations in time and
only in the spatial variable $\phi$, while both the axial as well
as the radial coordinates are parameters. It should be stressed
that this is not an extra approximation but rather follows from
the special geometry under investigation. In particular, it is
noticed that that sub-set does not contain any pressure effects.
The absence of pressure gradients in the horizontal dynamics hints
also to the absence of the SRI in thin discs. Thus, assuming the
following form for the solution of the sub-set
(\ref{13})-(\ref{14}):
 \begin{equation}
\{v_r', v_\phi'\}=\exp(i\omega t-i m\phi)\{\hat{v}_r,
\hat{v}_\phi\}+c.c.,\,\,\,\,\,\,\,\,\,\,\,\,\,\,\,\,\big{(}v_r'=\frac{\bar{n}V_r'}{\Omega},
v_\phi'=\frac{\bar{n}V_\phi'}{\Omega}\big{)},
 \label{19}
\end{equation}
where $m=0,\pm 1,\pm 2,\ldots$  is the azimuthal wave number,
$\omega$ is the frequency of perturbations. The reason for
transforming to the scaled mass flux variables defined in Eqs.
(\ref{19}) is that, as will be shown below, most of the resulting
equations do not depend on the radial coordinate. The
corresponding dispersion relation is given by:
 \begin{equation}
 \omega=m \Omega(r) \pm \chi(r).
 \label{20}
\end{equation}
This represents, not surprisingly, the stable epicyclic
oscillations in the disc plane. As $r$  and $\zeta$ are mere
parameters each  ring $r\equiv const$, $\zeta\equiv const$
vibrates independently in its own plane. If viscous stresses are
taken into account, an axial profile is imposed due to the mutual
shear stresses between the rings and each entire cylindrical shell
vibrates independently. This is indeed the case that has been
solved in [\cite{Umurhan et al. 2006} and \cite{Rebusco et al.
2009}].
Below  $\chi=\Omega(r)$  for the Keplerian rotation will be used
for simplicity.

As the axial and radial coordinates play the role of passive
parameters, the eigenfunctions of the inertia-Coriolis modes are
determined up to arbitrary amplitude $A(r,\zeta)$:
\begin{equation}
\hat{v}_r(r,\zeta)=A(r,\zeta), \hat{v}_\phi(r,\zeta)=\pm i
\frac{1}{2} A(r,\zeta).\,\,\,\,
 \label{21}
\end{equation}
As an example the following form of the amplitudes is considered
\begin{equation}
A(r,\zeta)=F(r)G(\zeta),
 \label{22}
\end{equation}
where the  $F(r)$ and $G(\zeta)$  are arbitrary functions. Any
special form of functions $A(r,\zeta)$,  or  $F(r)$ and $G(\zeta)$
represent some given set of  initial conditions. That special form
of the amplitude may be utilized to satisfy various regularity
conditions of the solutions near the disc's edges. The regularity
problem will be addressed in more details in sections 4 and 5.

Finally, it is of interest to follow the evolution in time of the
radial dependence of the perturbations. Thus, defining a
generalized radial wave number by, say, $k_r=i\partial ln
V_r'/\partial r$, it is readily seen that it depends on time as
$k_r=k_r^0-t\lambda d\Omega/ dr$ with the constant $\lambda$.
Thus, even initial radially uniform perturbation will eventually
develop strong radial dependence due to the rotation shear [see
also \cite{Goldreich and Lynden-Bell}, and \cite{Balbus and Hawley
1991}]. As a result, in a general case an initial value problem
has to be solved rather than a spectral one. In the current
supersonically rotating thin disc case, however, due to the
absence of radial partial derivative from the linearized
sub-system (which is due to the insignificant role played by the
radial steady-state as well as perturbed pressure gradients) a
spectral problem may be solved in which the generalized time
dependent radial wave number may be inferred from the solution.

\subsection{Vertical sound waves}
For frequencies that are different from  $\omega=m \Omega(r) \pm
\chi(r)$ in  (\ref{20}) both perturbed components of the
horizontal mass flux  are zero, i.e. $\hat{v}_r=0$ and
$\hat{v}_\phi=0$. As a result, Eqs. (\ref{15}) and (\ref{16})
describe the perturbations in the axial velocity and the number
(or mass) density. Since  $\hat{v}_r=0$  the radial derivative is
once again dropped from the problem (from Eq. (\ref{16})). Thus,
the following normal-wave solution is assumed:
 \begin{equation}
\{ v_z',n'\}=\exp(i\omega t-i m\phi)\{\hat{v}_z,
\hat{n}\}+c.c.,\,\,\,\,\,\,\,\,\,\,\,\,\,\,\,\,\big{(}v_z'=\frac{\bar{n}V_z'}{\Omega}\big{)}.
 \label{23}
\end{equation}
Writing  $\omega=\lambda \Omega(r)$ where $\lambda$  is a constant
(an assumption that will be justified a posteriori) and
substituting (\ref{21}) - (\ref{23}) into (\ref{15}) -(\ref{16})
yields:
\begin{equation}
i(\lambda-m)\hat{v}_z - \frac{2-\gamma}{\gamma-1}
\frac{d \bar{c}_S^2}{d \zeta}\hat{n}+\bar{c}_S^2(\zeta)
\frac{\partial \hat{n}}{\partial \zeta}=0,
 \label{24}
\end{equation}
\begin{equation}
i(\lambda-m)\hat{n}+\frac{\partial \hat{v}_z}{\partial \zeta}=0.
 \label{25}
\end{equation}
%

Finally, Eqs. (\ref{24}) and (\ref{25})
result in the following second order ordinary
differential equation: 
\begin{equation}
(1-\zeta^2)\frac{\partial^2 \hat{n}}{\partial \zeta^2}-
\frac{2\gamma-3}{\gamma-1}2\zeta
\frac{\partial \hat{n}}{\partial \zeta}+\alpha^2\hat{n}=0,\,\,\,\,
\alpha^2=\frac{2}{\gamma-1}[(\lambda-m)^2+2-\gamma].
 \label{27}
\end{equation}
In general, the solutions of Eq. (\ref{27}), complemented by the
boundary conditions $\hat{n}=0$  at the disc edges $\zeta=\pm 1$,
may be expressed in terms of hypergeometric functions. However,
for $\gamma=5/3$ the solutions become remarkably simple and may be
expressed in terms of the following symmetric and anti-symmetric
eigenfunctions (Appendix A):
\begin{equation}
\alpha_k=2k+1,\,\,\,\,\hat{n}=N_{2k+1}\cos(2k+1)\theta,\,\,\,\,
k=0, \pm 1, \pm 2,\ldots,
 \label{28}
\end{equation}
\begin{equation}
\alpha_k=2k,\,\,\,\,\hat{n}=N_{2k}\sin 2k\theta,\,\,\,\,
k=\pm 1, \pm 2,\ldots,
 \label{29}
\end{equation}
where  $\theta=\arcsin \zeta$. Thus, according to (\ref{27}) the
following dispersion relation is obtained for  $\gamma=5/3$:
\begin{equation}
\lambda=\lambda_\pm^{(m,k)}=m\pm
\sqrt{\frac{\alpha^2_k-1}{3}},\,\,\,m=0,\pm 1, \pm 2,\ldots.
 \label{30}
\end{equation}
The admissible values of  $\alpha_k$ and $k$  are given by either
(\ref{28}) or (\ref{29}). These eigenvalues describe stable
stratified sound waves that propagate independently on the
surfaces of each individual cylindrical shell. In a co-rotating
frame those are standing sound waves whose dependence on the axial
coordinate is described by Eqs. (\ref{28}) or (\ref{29}), where
$k$ plays the role of an axial wave number. Thus, according to
(\ref{28}) or (\ref{29}) as $k$ is increased the waves concentrate
in growing numbers towards the disc's edges. It is finally
emphasized that the compressibility effects in the vertical
direction are indeed essential for the thin disc approximation as
the sound vertical crossing time is of the order of $\Omega^{-1}$.

The seemingly simple dispersion relations (\ref{20}) and
(\ref{30}) actually entail an important conclusion. As they
exhaust all possible eigenvalues of the linearized system it means
that in the limit of thin disc approximation, the disc is
spectrally stable against hydrodynamic perturbations. Thus, the
asymptotic in time behavior of the linearized system exhibit two
stable families of normal mode oscillations within the thin discs;
one consists of vibrating rings due to inertia-coriolis effects,
while the other one imposes density fluctuation on each individual
cylindrical shell in the form of standing sound waves. In
particular, it is obvious that no unstable strato-rotational modes
may be excited in such systems. This is due to the absence of
radial and azimuthal perturbed pressure gradients from Eqs.
(\ref{13}) and (\ref{14}), which leads to the decoupling of the
vertical density (strato) from the planar (rotational) dynamics.
In contrast to that picture, in Couette flows, such pressure
gradients play an important role in coupling of the planar and
vertical perturbations and thus giving rise to the
strato-rotational instability. The reason for the importance of
the planar pressure gradients in Couette flows is the presence of
a rigid wall, which is absent in rotating discs in thin disc
approximation.

Having clarified the time asymptotic (marginal) stability of
rotating discs, this is by no means the end of the hydrodynamical
story. Short time dynamics may exhibit significant amplification
of initial small perturbations. As  will be shown below, two
mechanisms are responsible for such behavior, namely resonant as
well as non-resonant excitation of vertical sound waves by planar
inertia-coriolis modes. Furthermore, the algebraically growing
perturbations can create the conditions for the re-appearing of
the SRIs. It should be mentioned that \cite{Umurhan et al. 2006}
and \cite{Rebusco et al. 2009} have described the non resonant
mechanism in discs that are characterized by a phenomenological
viscosity, by expansion in a single small parameter. The current
work explores asymptotic expansions in the small but finite aspect
ratio of the steady-state disc,  $\epsilon$, for both equilibrium
and perturbations, while the amplitude of the perturbations is
assumed to be infinitesimal small independent value. The resonant
mechanism is described here for the first time.

\section{Non-resonantly driven sound waves}
Returning to Eqs. (\ref{15}) and (\ref{16}) that describe the
dynamics of the vertical sound waves, it is clear that the latter
may be excited by the inertia-coriolis modes that are eigen
functions of the decoupled system (\ref{13})-(\ref{14}). Moreover,
due to the rotation shear, the amplitude of the driven sound modes
grows linearly in time. This is illustrated by Eq. (\ref{16})
which may be cast with the aid of expressions
(\ref{19})-(\ref{22}) in the following form of the auxiliary
inhomogeneous equation:
\begin{equation}
\frac{1}{\Omega}\frac{\partial n'}{\partial t}+ \frac{\partial n'
}{
\partial \phi } + \frac{\partial v_z'}{\partial z }=
-\big{(}i t \lambda
\frac{d\Omega}{dr}+\frac{1}{F}\frac{dF}{dr}+\frac{1}{r\Omega}\frac{dr\Omega}{dr}\big{)}v_r'
+i\frac{m}{r}v_\phi'.
\,\,\,\,\,\,\,\,\,\,\,\,\,\,\,\,\,\,\,\,\,\,\,\,\,\,\,\,\,
\,\,\,\,\,\,\,\,\,\,\,\,\,\,\,\,\,\,\,\,\,\,\,\,\,\,\,\,\,\,\,\,
\,\,\,\,\,\,\,\,\,\,\,\,\,\,\,\,\,\,\,\,\,\,\,\,\,\,\,\,\,\,\,\,\,\,\,\,
\,\,\,\,\,\,\,\,\,\,\,\,\,\,\,\,\,\,\,\,\,\,\,\,\,\,\,\,\,\,\,\,\,\,
\,\,\,\,\,\,\,\,\,\,\,\,\,\,\,\,\,\,\,\,\,\,\,\,\,\,\,
\label{coupling}
\end{equation}

%
The solutions of the inhomogeneous Eqs. (\ref{15})-(\ref{16}) are
represented now in the following way:
 \begin{equation}
\{v_r',v_\phi', v_z',n'\}=\exp(i\omega t-i
m\phi)\{\hat{v}_r,\hat{v}_\phi,\hat{v}_z,\hat{n}\}+c.c.
 \label{31}
\end{equation}
Here  $\omega=\lambda \Omega(r)$ ($\lambda \equiv const$)
   represents the eigen-frequency of the inertia-coriolis
waves, i.e., $\lambda=m \pm 1$, ($m=0, \pm 1, \pm 2,\ldots$)
according to Eq. (\ref{20}). In addition it is assumed that
 \begin{equation}
\hat{v}_r=\hat{v}_r^{(0)},\hat{v}_\phi=\hat{v}_\phi^{(0)},
\hat{v}_z=\hat{v}_z^{(0)}+it\Omega\hat{v}_z^{(1)},\hat{n}=\hat{n}^{(0)}+it\Omega\hat{n}^{(1)}.
 \label{32}
\end{equation}
Here the superscript denotes the order in time; $\hat{v}_r^{(0)}$,
$\hat{v}_\phi^{(0)}$ are given by Eqs. (\ref{21}),(\ref{22}) while
relations for $\hat{n}^{(k)}$  and $\hat{v}_z^{(k)}$  ($k=0,1$)
are derived after some algebra by substituting (\ref{31}) and
(\ref{32}) into (\ref{15}) - (\ref{16}) and equating equal powers
of $t$:
 \begin{equation}
\pm i\hat{n}^{(1)}+\frac{\partial \hat{v}_z^{(1)}}{\partial
\zeta}=-\lambda \frac{1}{\Omega}\frac{d\Omega}{d r}\hat{v}_r,
 \label{33}
\end{equation}
\begin{equation}
\pm i\hat{v}_z^{(1)}- \frac{2-\gamma}{\gamma-1}
\frac{d \bar{c}_S^2}{d \zeta}\hat{n}^{(1)}+\bar{c}_S^2(\zeta)
\frac{\partial \hat{n}^{(1)}}{\partial \zeta}=0,
 \label{34}
\end{equation}
and
 \begin{equation}
\pm i\hat{n}^{(0)}+\frac{\partial \hat{v}_z^{(0)}}{\partial
\zeta}=-i\hat{n}^{(1)}
- \frac{1}{r\Omega}\frac{\partial r\Omega \hat{v}_r}{\partial r}
+\frac{im}{r}  \hat{v}_\phi,
 \label{35}
\end{equation}
\begin{equation}
\pm i\hat{v}_z^{(0)}- \frac{2-\gamma}{\gamma-1}
\frac{d \bar{c}_S^2}{d \zeta}\hat{n}^{(0)}+\bar{c}_S^2(\zeta)
\frac{\partial \hat{n}^{(0)}}{\partial \zeta}=-i\hat{v}_z^{(1)},
 \label{36}
\end{equation}
with the following boundary conditions:
\begin{equation}
\hat{n}^{(0)}=\hat{n}^{(1)}=0\,\,\,\mbox{for}\,\,\,\zeta=\pm 1.
 \label{37}
\end{equation}

As is indeed seen from Eqs. (\ref{33}) -  (\ref{36})  the sound
waves are driven by the inertial mode, while the linear growth in
time is entirely due to effect of the rotation shear. Substituting
now Eqs.  (\ref{21}), (\ref{22})   into Eqs. (\ref{33}),
(\ref{34}) and  (\ref{35}),  (\ref{36}) yields the following two
decoupled subsystems:
\begin{equation}
(1-\zeta^2)\frac{\partial^2 \hat{n}^{(1)}}{\partial \zeta^2}-
\frac{2\gamma-3}{\gamma-1}2\zeta
\frac{\partial \hat{n}^{(1)}}{\partial
\zeta}+2\frac{3-\gamma}{\gamma-1}\hat{n}^{(1)}=\pm 2i\frac{m\pm
1}{\gamma-1}\frac{F(r)}{\Omega(r)}\frac{d\Omega}{dr}G(\zeta),\,\,\,\,
 \label{38}
\end{equation}
\begin{equation}
\hat{v}_z^{(1)}=\pm i (2-\gamma)\zeta
\hat{n}^{(1)}\pm i\frac{\gamma-1}{2}(1-\zeta^2)\frac{\partial
\hat{n}^{(1)}}{\partial \zeta},
 \label{39}
\end{equation}
and
\begin{equation}
(1-\zeta^2)\frac{\partial^2 \hat{n}^{(0)}}{\partial \zeta^2}-
\frac{2\gamma-3}{\gamma-1}2\zeta
\frac{\partial \hat{n}^{(0)}}{\partial
\zeta}+2\frac{3-\gamma}{\gamma-1}\hat{n}^{(0)}=
\mp \frac{4}{\gamma-1}\hat{n}^{(1)}
+i\frac{1}{\gamma-1}\big{[} 2(m\pm
1)\frac{F}{\Omega}\frac{d\Omega}{dr}+i m\frac{F}{r}
\pm \frac{2}{r\Omega}\frac{d r F\Omega}{dr} \big{]}G,
 \label{40}
\end{equation}
\begin{equation}
\hat{v}_z^{(0)}=\pm i (2-\gamma)\zeta
\hat{n}^{(0)}\pm i\frac{\gamma-1}{2}(1-\zeta^2)\frac{\partial
\hat{n}^{(0)}}{\partial \zeta}\mp\hat{v}_z^{(1)}.
 \label{41}
\end{equation}

Equations (\ref{38}) and (\ref{40}), subject to the boundary
conditions (\ref{37}) and either symmetric or anti-symmetric
scaling function $G(\zeta)$ are consistent with either symmetric
or anti-symmetric solutions $\hat{n}^{(k)}(\zeta)$ ($k=0,1$), i.e:
\begin{equation}
\hat{n}^{(k)}(\zeta)=\hat{n}^{(k)}(-\zeta),
 \label{42}
\end{equation}
\begin{equation}
\hat{n}^{(k)}(\zeta)=-\hat{n}^{(k)}(-\zeta).
 \label{43}
\end{equation}

A particular family of solutions is considered which is
characterized by the following form of the free functions $F(r)$
and  $G(\zeta)$ for  $\gamma=5/3$  (Appendix A):
\begin{equation}
F(r)=r,\,\,\,\,G(\zeta)=G_3\cos^3\theta+G_5\cos^5\theta+G_7\cos^7\theta,\,\,\,\,(\theta=\arcsin\zeta).
 \label{44}
\end{equation}
As is shown in Appendix A3, such choice of $F(r)$ and  $G(\zeta)$
results in radial independence of the perturbed number density and
its regularity at the disc edges. In that case the problem is
reduced to two second order ordinary differential equations for
$\hat{n}^{(0)}$ and $\hat{n}^{(1)}$  complemented by the
corresponding boundary conditions (\ref{42}) or (\ref{43}) at the
disc edges, while $\hat{v}_z^{(0)}$  and $\hat{v}_z^{(1)}$ may be
calculated from (\ref{39}) and (\ref{41}). Thus, using (\ref{44})
the following symmetric solutions for $\hat{n}^{(0)}$ and
$\hat{n}^{(1)}$ are obtained:
\begin{equation}
\hat{n}^{(k)}(\zeta)=N_3^{(k)}\cos^3\theta+N_5^{(k)}\cos^5\theta+N_7^{(k)}\cos^7\theta,
 \label{45}
\end{equation}
where the coefficients $N_j^{(k)}$  and  $G_j$ ($j=3,5,7$,
$k=0,1$) are independent of the radius. Thus, for each value $m$
there are two linearly growing modes except for $m=\pm1$, where
for each $m$ corresponds only a single growing mode. It is finally
noted that by the special construction of the solution in Appendix
A3 both right hand sides of Eqs. (\ref{38}) and (\ref{40}) satisfy
the solvability condition, namely that they are orthogonal to the
solution of the common homogeneous part.

The special choice of the free function  $G(\zeta)$ presented in
Eq. (\ref{44})  indeed assures the regularity of all components of
the velocity vector $\bold{V}'$. However, such regularity
requirements are rather too stringent and actually the sharp
boundary model used in the current work does allow certain degree
of singularity in the velocity. Thus, more relaxed conditions that
are easily satisfied are that the perturbed components of the mass
flux (or the momentum per unit volume), i.e.  $\bar{n}\bold{V}'$
is zero at the disc's edges. Indeed, such singularity in the
velocity components is a direct consequence of the sharp boundary
disc with zero equilibrium-density at the horizontal disc edges
($\zeta = \pm 1$) that has been employed in the present
investigation. In fact, astrophysical discs have no well defined
boundaries, the disc boundary is not sharp, and the vacuum
boundary conditions imposed at a finite height are rather a good
approximation than an exact model. According to a more realistic
model that assumes a diffused disc with exponentially decreasing
density [see e.g. \cite{Balbus and Hawley 1991} for thin
isothermal discs], the density has a small but finite value at the
effective boundary of the disc. With that in mind the singularity
at the disc edges may be regularized within a model with a small
but non zero pressure value at the effective disc edges. The
regularized model yields large but finite values of the fluid
velocities at the disc edge that demonstrate that non modal
growth may indeed leads to significant amplification of the
perturbations.

An example of a singular case is:
\begin{equation}
F(r)=r,\,\,\,\,G(\zeta)=G_1\cos\theta,\,\,\,\,(\theta=\arcsin\zeta).
 \label{46}
\end{equation}
Again such choice of $F(r)$  provides the solutions for
$\hat{n}^{(k)}$ ($k=0,1$) which are independent of the  radius:
\begin{equation}
\hat{n}^{(k)}=N_1^{(k)}\cos\theta, \,\,\,\,
 (N_1^{(1)}=\mp i
\frac{3}{2}(m\pm1)G_1, \,\,\,\,
N_1^{(0)}= - i (4m\pm 5)G_1).
 \label{47}
\end{equation}
In this case the velocity indeed has singularity at the disc's
edges, however due to the more rapid decrease of the steady state
number density the mass flux is zero.

\section{ Resonantly driven sound waves}
As can be seen from Eq. (\ref{coupling}), stable sound modes may
be excited resonantly by the inertial modes if any pair of
respective eigen-values (\ref{20})
 (with
$\chi=\Omega(r)$, $\omega=\lambda_{\pm}^{(m)}\Omega(r)$, where
$\lambda_{\pm}^{(m)}=m\pm 1$)
and (\ref{30}) are equal, i.e.
$\lambda_{\pm}^{(m)}=\lambda_{\pm}^{(m,k)}$ ($m=0,\pm 1, \pm 2,
\ldots$, $k=0,1$). It is easy to show that this happens only for
the first anti-symmetric mode (\ref{29}) for arbitrary $m$
($\gamma=5/3$) and:
 \begin{equation}
k=\pm 1,\,\,\,\alpha_k=\pm 2,\,\,\,\lambda_{\pm}^{(m,k)}=m\pm
1,\,\,\,
\hat{n}=\pm N_2\sin2\theta,\,\,\,\,(\theta=\arcsin\zeta).
 \label{48}
\end{equation}

The solution of Eq. (\ref{coupling}) then, that describes the
resonant coupling between sound and inertial waves in Keplerian
discs is thus represented by:
  \begin{equation}
\{v_r',v_\phi', v_z', n'\}=\exp(i\omega t-i
m\phi)\{\hat{v}_r,\hat{v}_\phi,\hat{v}_z,\hat{n}\}+c.c.,\,\,\,\omega=\lambda
\Omega(r),\,\,\, (\lambda \equiv const).
 \label{49}
\end{equation}
The combination of resonant coupling with the effect of the
rotation shear leads to the following form of the solution:
 \begin{equation}
\hat{v}_r=\hat{v}_r^{(0)},\,\,\hat{v}_\phi=\hat{v}_\phi^{(0)},\,\,
\hat{v}_z=\hat{v}_z^{(0)}+it\Omega\hat{v}_z^{(1)}+\frac{(it\Omega)^2}{2}\hat{v}_z^{(2)},\,\,
\hat{n}=\hat{n}^{(0)}+it\Omega\hat{n}^{(1)}+\frac{(it\Omega)^2}{2}\hat{n}^{(2)},
 \label{50}
\end{equation}
where the superscripts denote the order in time;
$\hat{v}_r^{(0)}$,  $\hat{v}_\phi^{(0)}$ are given by  (\ref{21}),
 (\ref{22}) while  $\hat{v}_z^{(k)}$, $\hat{n}^{(k)}$, ($k=0,1,2$) are determined by the relations below. It
should be noted that resonant interaction between the two modes of
wave propagation may occur also in a shearless rotation. In that
case the amplitude of the sound waves grows only linearly in time.
It is the combined effect of resonant interaction and rotation
shear that gives rise to a quadratic growth in time that is
described in Eq. (\ref{50}). Substituting (\ref{49})- (\ref{50})
into (\ref{13})) - (\ref{16}) and equating equal powers of $t$
yields
 \begin{equation}
\pm i\hat{n}^{(2)}+\frac{\partial \hat{v}_z^{(2)}}{\partial
\zeta}=0,
 \label{51}
\end{equation}
\begin{equation}
\pm i\hat{v}_z^{(2)}- \frac{2-\gamma}{\gamma-1}
\frac{d \bar{c}_S^2}{d \zeta}\hat{n}^{(2)}+\bar{c}_S^2(\zeta)
\frac{\partial \hat{n}^{(2)}}{\partial \zeta}=0,
 \label{52}
\end{equation}
and
 \begin{equation}
\pm i\hat{n}^{(1)}+\frac{\partial \hat{v}_z^{(1)}}{\partial
\zeta}=-i\hat{n}^{(2)}
- \lambda\frac{1}{\Omega}\frac{d\Omega }{d r}\hat{v}_r
%
 \label{53}
\end{equation}
\begin{equation}
\pm i\hat{v}_z^{(1)}- \frac{2-\gamma}{\gamma-1}
\frac{d \bar{c}_S^2}{d \zeta}\hat{n}^{(1)}+\bar{c}_S^2(\zeta)
\frac{\partial \hat{n}^{(1)}}{\partial \zeta}=0,
 \label{54}
\end{equation}
and
 \begin{equation}
\pm i\hat{n}^{(0)}+\frac{\partial \hat{v}_z^{(0)}}{\partial
\zeta}=-i\hat{n}^{(1)}
- \frac{1}{r\Omega}\frac{\partial r\Omega \hat{v}_r}{\partial r}
+\frac{im}{r}  \hat{v}_\phi,
 \label{55}
\end{equation}
\begin{equation}
\pm i\hat{v}_z^{(0)}- \frac{2-\gamma}{\gamma-1}
\frac{d \bar{c}_S^2}{d \zeta}\hat{n}^{(0)}+\bar{c}_S^2(\zeta)
\frac{\partial \hat{n}^{(0)}}{\partial \zeta}=-i\hat{v}_z^{(1)},
 \label{56}
\end{equation}
while the planar dynamics is described by Eqs. (\ref{21}),
(\ref{22}) for $\hat{v}_r$ and   $\hat{v}_\phi$. The number
densities $\hat{n}^{(k)}$ ($k=0, 1, 2$) are supplemented by the
boundary conditions:
\begin{equation}
\hat{n}^{(0)}=\hat{n}^{(1)}=\hat{n}^{(2)}=0\,\,\,\mbox{for}
\,\,\,\zeta=\pm 1.
 \label{57}
\end{equation}
Equations (\ref{55}) and (\ref{56}) for  $\hat{n}^{(2)}$ and
$\hat{v}_z^{(2)}$ form a homogeneous sub-system for the sound
mode:
\begin{equation}
(1-\zeta^2)\frac{\partial^2 \hat{n}^{(2)}}{\partial \zeta^2}-
\frac{2\gamma-3}{\gamma-1}2\zeta
\frac{\partial \hat{n}^{(2)}}{\partial
\zeta}+2\frac{3-\gamma}{\gamma-1}\hat{n}^{(2)}=0,\,\,\,\,
 \label{58}
\end{equation}
\begin{equation}
\hat{v}_z^{(2)}=\pm i (2-\gamma)\zeta
\hat{n}^{(2)}\pm i\frac{\gamma-1}{2}(1-\zeta^2)\frac{\partial
\hat{n}^{(2)}}{\partial \zeta}.
 \label{59}
\end{equation}
The only solutions of Eq. (\ref{58}) that satisfy the resonance
conditions and which are zero at $\zeta=\pm 1$ and satisfy
(\ref{57}) are given by ($\gamma=5/3$):
 \begin{equation}
\hat{n}^{(2)}=\pm N_2^{(2)}\sin2\theta,\,\,\,
\hat{v}_z^{(2)}=iN_2^{(2)}
[(2-\gamma)\sin\theta \sin2\theta+(\gamma-1)\cos\theta
\cos2\theta],
\,\,\,(\theta=\arcsin\zeta).
 \label{60}
\end{equation}
In general, $N_2^{(2)}$  may depend on the radial variable,
however, as an example it will be considered as a constant. Also,
it is noted that in contrast to the non-resonant case, here
singularity in the velocity is inevitable. However, as stated
above, the less restrictive condition is satisfied as the mass
flux is regular and vanishes at the disc's edges.

Moving on the following equations for  $\hat{n}^{(1)}$,
$\hat{v}_z^{(1)}$  and  $\hat{n}^{(0)}$, $\hat{v}_z^{(0)}$:
\begin{equation}
(1-\zeta^2)\frac{d^2 \hat{n}^{(1)}}{d \zeta^2}-
\frac{2\gamma-3}{\gamma-1}2\zeta
\frac{d\hat{n}^{(1)}}{d
\zeta}+2\frac{3-\gamma}{\gamma-1}\hat{n}^{(1)}=\mp\frac{2}{\gamma-1}\hat{n}^{(2)}
\mp 3i\frac{m\pm 1}{\gamma-1}G(\zeta),\,\,\,\,
 \label{62}
\end{equation}
\begin{equation}
\hat{v}_z^{(1)}=\pm i (2-\gamma)\zeta
\hat{n}^{(1)}\pm i\frac{\gamma-1}{2}(1-\zeta^2)\frac{d
\hat{n}^{(1)}}{d \zeta},
 \label{63}
\end{equation}
and
\begin{equation}
(1-\zeta^2)\frac{d^2 \hat{n}^{(0)}}{d \zeta^2}-
\frac{2\gamma-3}{\gamma-1}2\zeta
\frac{d \hat{n}^{(0)}}{d
\zeta}+2\frac{3-\gamma}{\gamma-1}\hat{n}^{(0)}=
- \frac{2}{\gamma-1}\hat{n}^{(2)} 
\mp \frac{4}{\gamma-1}\hat{n}^{(1)}
-2i\frac{m\pm 1}{\gamma-1}G,
 \label{64}
\end{equation}
\begin{equation}
\hat{v}_z^{(0)}=\pm i (2-\gamma)\zeta
\hat{n}^{(0)}\pm i\frac{\gamma-1}{2}(1-\zeta^2)\frac{d
\hat{n}^{(0)}}{d \zeta}\mp\hat{v}_z^{(1)}.
 \label{65}
\end{equation}

The homogeneous part of Eq. (\ref{62}) complemented by
corresponding boundary conditions for $\hat{n}^{(1)}$ coincides
with the above eigenvalue problem for $\hat{n}^{(2)}$. Hence
non-trivial solutions of the inhomogeneous equations exist if the
right hand sides of  (\ref{62}) satisfies the solvability
condition, namely, that it is orthogonal to $\hat{n}_2$ as it is
given in (\ref{60}). This means that in order to obtain resonant
coupling, the vertical dependence of the inertia-coriolis waves
must contain a term of the following shape:
\begin{equation}
G(\zeta)=G_2 \sin2\theta,\,\,\,(\theta= \arcsin \zeta),
 \label{61}
\end{equation}
where $G_2$  is an arbitrary amplitude. For simplicity it will be
assumed that Eq. (\ref{61}) indeed describes the entire profile of
the driving waves and in addition that $F(r)=r$. As will be seen
later, the latter results in simple radial independent solutions.
Imposing now the solvability condition, $N_2^{(2)}$ is given by:
\begin{equation}
N_2^{(2)}=\mp i\frac{3}{2}(m\pm 1)G_2 .
 \label{66}
\end{equation}

Eq. (\ref{62}) for $\hat{n}^{(1)}$ may be solved now. Due to
imposing the solvability condition and the special form of $G$
given in Eq. (\ref{61}), the right hand side of that equation is
zero. Hence its solution is given by $N_2^{(1)}\sin2 \theta$.
Turning now to Eq. (\ref{64}) it is immediately seen that the same
solvability condition determines now $N_2^{(1)}$ to be:
\begin{equation}
N_2^{(1)}=\pm i\frac{1}{4}(m\mp 1)G_2.
 \label{661}
\end{equation}

Summarizing, the solutions for $\hat{n}^{(1)}$,  $\hat{v}_z^{(1)}$
and $\hat{n}^{(0)}$, $\hat{v}_z^{(0)}$  are
($\theta=\arcsin\zeta$):
\begin{equation}
\hat{n}^{(1)}= N_2^{(1)}\sin2\theta,\,\,\,
\hat{v}_z^{(1)}=\pm iN_2^{(1)}
[(2-\gamma)\sin\theta \sin2\theta+(\gamma-1)\cos\theta
\cos2\theta],
%
 \label{67}
\end{equation}
\begin{equation}
\hat{n}^{(0)}= N_2^{(0)}\sin2\theta,\,\,\,
\hat{v}_z^{(0)}= i[\pm N_2^{(0)}- N_2^{(1)}\Omega(r)]
[(2-\gamma)\sin\theta
\sin2\theta+(\gamma-1)\cos\theta\cos2\theta].
%
 \label{68}
\end{equation}
In the above solution (\ref{66}) - (\ref{68}) there are two
arbitrary constants $G_2$  and  $N_2^{(0)}$. The first one
represents the arbitrary amplitude of the driving inertia-coriolis
waves while the second one describes free sound oscillations.
Finally, it is noticed again that while the vertical component of
the perturbed velocity is singular at the disc's edges, the
corresponding component of the mass flux  tends to zero.

\section{ Conclusions}
A comprehensive study of the long as well as short time
hydrodynamic response of Keplerian rotating discs to small
perturbations has been carried out for realistic thin disc
geometry. Explicit analytical solutions of the classical
eigenvalue problem reveal that thin discs are hydrodynamically
marginally stable asymptotically in time. Such a picture naturally
leads to the ruling out of the strato-rotational instabilities as
primary contributors to the dynamical evolution of thin
ultrasonically rotating discs. This has been shown to be a direct
result of the decoupling of the planar dynamics from the vertical
acoustics, which in turn is a result of the negligible role played
by radial pressure gradients. The latter is indeed the agent of
coupling of the strato and rotational modes that lead to
instability in the case of Couette flows due to the presence of
solid radial boundaries.  Notwithstanding the spectral stability
of thin rotating discs, they have been shown to exhibit intense
hydrodynamical activity in the form of algebraic short-term growth
of initially small perturbations. Two mechanisms have been
identified and studied that lead to such non modal dynamical
behavior. The first one is the non resonant excitation of
co-rotating standing sound waves on individual cylindrical shells
by planar inertia-coriolis modes. Such coupling between the modes
occurs due to the rotation shear and results in a linear in time
growth of the perturbations. The second mechanism is the resonant
driving of the sound waves by the inertia-coriolis modes. While
non resonant coupling relies on the non normal nature of the
linearized system of the dynamical equations, resonant coupling
exists also for a shearless rotation.  However, in the presence of
rotation shear the perturbations grow quadratically in time. Both
mechanisms are powered by the compressible motion of the fluid.
Thus, compressible inertia-coriolis stable oscillations
continuously pump energy from the sheared steady state flow and
transfer it (resonantly as well as non resonantly) to the
continuously algebraically growing sound waves. Compressibility is
indeed inevitable due to the supersonic rotation of the
steady-state disc combined with its small vertical dimensions
which make the vertical sound crossing time of the order of a
rotation period. Simple explicit analytical solutions were
obtained for all possible cases. It should be stressed that non
modal growth has been exhibited for all admissible parameters of
the system.

Along side with the non modal growth of small perturbations, the
rotation shear is also responsible for the linear growth in time
of an effective radial wave number. Such a process naturally gives
rise to enhanced perturbed radial pressure gradients. In light of
the discussion of the special role played by the latter in the
onset of the SRI, it may be speculated, that such a development
may reinstate the SRI as an important dynamical process due to
secondary instabilities that occur in the wake of the primary non
modal hydrodynamical growth. Thus, the linear as well as quadratic
amplification of initial small perturbation as well as the
secondary instabilities they entail could provide a viable
mechanism for sustaining turbulence and consequently to enhanced
transport coefficients.

A detailed description of nonmodal growth has been presented
before in \cite{Umurhan et al. 2006} and \cite{Rebusco et al.
2009}. The main differences between those works and the current
investigation are:
\begin{itemize} \item In \cite{Umurhan et al.
2006} and \cite{Rebusco et al. 2009} a finite amplitude non-linear
solution to the problem of the dynamical evolution of the full
equations was sought. In that approach it is assumed that
solutions of the equations, both steady as well as dynamical, may
be expanded in a single small parameter, namely, $\epsilon$.
Solutions are then determined by iteratively solving the resulting
equations at each order of $\epsilon$. Consequently, the forcing
terms at each order are the solutions of the previous order which
results in algebraic growth. In contrast, in the current work a
classical linearization procedure is employed in which an extra
independent small parameter is introduced that measures the
amplitude of the initial perturbations. It is due to the
negligible role played by the radial pressure gradients that the
resulting linear system of equations is decoupled into two
subsystems, one of which describes the perturbed planar velocities
while the other one determines the dynamical evolution of the
vertical sound waves. Moreover, the planar modes act as forcing
terms for the vertical sound waves. This mechanism gives rise to
the resonant as well as the nonresonant algebraic growth of the
latter. \item \cite{Umurhan et al. 2006} and \cite{Rebusco et al.
2009} have demonstrated the occurrence of linear
growth
(due to the shear of the steady state rotation)
that results from the nonresonant driving of the vertical sound
waves by the planar dynamics. In the current work, in addition to
that linear growth, a resonant process is identified which results
in a quadratic growth in time. Such second order growth is a
result of the combined effect of the resonance between the
vertical and and the planar modes and the shear of the steady
state rotation. Furthermore, the resonance is a global one in the
sense that it occurs all over the disk; the planar vibrations of a
ring of any radius resonantly rattle the cylindrical shell where
it resides. \item Unlike the inviscid limit that is investigated
in the current work, \cite{Umurhan et al. 2006} and \cite{Rebusco
et al. 2009} present the solution of the viscous problem where the
viscous stresses are modelled by the Shakura-Sunyaev $\alpha$
parameter. \cite{Umurhan et al. 2006} and \cite{Rebusco et al.
2009} refer to values around $\alpha = 10^{-3}$ as representing
"realistic" values as they are close to the result found
numerically in subcritical hydrodynamic transitions in \cite{Lesur
and Longaretti 2005}. Numerical solutions indicate that for such
small values of the viscosity (or large values of the Reynold
number), the algebraic growth proceeds undisturbed at least for
$10^3$ rotation times, during which the amplitude grows by a
factor of $10^4$, before viscous damping starts to play any
significant role. Thus, the dynamical evolution during the first
non-viscous long period of time is independent of the viscosity
and is well described by the inviscid limit. Indeed, the
eigenvalues of the planar dynamics [Eq. (63) in \cite{Rebusco et
al. 2009}] as well as the temporal behavior of the driven sound
waves [Eq. (66) in \cite{Rebusco et al. 2009}] transit to Eq.
(\ref{20}) and Eqs. (\ref{31}) Eqs. (\ref{32}), respectively of
the current work in the limit $\alpha \rightarrow 0$. The singular
limit of zero viscosity is revealed then only in the spatial
profiles of the eigenfunctions. Thus, the spatial profiles
obtained in the current work may be considered as outer solutions
of a boundary layer problem which are valid in most parts of the
disk except perhaps within a thin region next to the vertical
edges of the disk. The advantage of the inviscid approach however
is that it enables to derive simple analytical expressions for the
eigenvalues and the eigenfunctions such as those given in Eqs.
(\ref{28})-(\ref{30}), and consequently to identify the resonant
quadratic growth. The inviscid problem has also been solved by
\cite{Umurhan and Shaviv 2005} where they have described the
nonresonant growth with a single parameter expansion (see first
item).

\end{itemize}

Obviously, the algebraically growing perturbations (both resonant
as well as non resonant) do not keep growing without bound. Thus,
nonlinear effects as well as viscous damping are expected to
quench the linear or quadratic growth of the perturbations. As has
been demonstrated in \cite{Rebusco et al. 2009}, for realistic
values of the viscosity, the algebraic growth of nonresonantly
driven sound waves takes place unhindered for at least $10^3$
rotation times before viscous damping kicks in. During that time
the initially small perturbations grow significantly by few orders
of magnitude. It is expected therefore, that nonlinear
redistribution of the perturbations energy among other length
scales will occur before the perturbations are damped by viscous
effects. In the case of resonantly driven sound waves the growth
of the perturbation is even faster. In addition, as $|k_r| \sim
|t\lambda d\Omega/dr|$
(see Section 3.1)
the time needed to develop significant perturbed radial pressure
gradients is of the order of $(\lambda \epsilon)^{-1}$. According
to \cite{Rebusco et al. 2009} there is enough time for that to
occur before viscous effects become important and even before
nonlinear effects take over. Moreover, the time to establish
significant radial gradients decreases as the azimuthal mode
number is increased. Also, according to \cite{Dubrulle et al.
2005} and \cite{Shalybkov and Rudiger 2005}, the maximal growth
rates of the SRI are achieved for Froude numbers of the order of
$\sim 1-10$ and are of the order of $10^{-1}$ inverse rotation
periods. This, again, leaves ample time for the growth of
secondary SRI instabilities before the perturbations are killed
off by viscous stresses. It should be mentioned however that in
\cite{Dubrulle et al. 2005} as well as in \cite{Shalybkov and
Rudiger 2005} (as indeed in most SRI investigations) the Froude
number has been assumed to be constant. This is very different
from realistic thin disk configurations where the Froude number is
given according to eqs. (\ref{10}) and (\ref{12}) by $Fr^2 =
3\zeta ^2 /(1-\zeta ^2)$ and thus varies between zero and infinity
over the small vertical extent of the disk. Hence, the numbers
cited above are used just as a crude estimation and the true
nature of a possible secondary instability may be investigated at
this stage only by numerical simulations.

It should finally be stressed once again that the calculations presented above are valid in
the Keplerian portion of the disk, namely far away from the radius of zero torque. In regions
around the latter, which are much closer to the central object, the radial pressure gradients
may play an important role in the dynamics of small perturbations, as is shown in \cite{Reynolds and
Miller 2009} who studied numerically g-mode trapping that occur within 8 - 32 Schwarzschild
radii from a central black hole.

\section*{Acknowledgments}
This work has been carried out while one of us (M.M) was a
Mercator Professor at the university of Potsdam. He would like to
thank the Physics department at the University of Potsdam and the
Astrophysikalisches Institut Potsdam for their hospitality. The
authors thank Georgy Burde for useful and insightful discussions.

\appendix
\section{}\label{app A}
\subsection{The basic boundary-value problem}
Equations (\ref{27}), (\ref{38}), (\ref{40}), (\ref{58}) and
(\ref{62}), (\ref{64}) have the following general form (it was
chosen $F(r)=r$ everywhere through that section):
\begin{equation}
(1-\zeta^2)\frac{d^2 n}{d \zeta^2}-
\frac{2\gamma-3}{\gamma-1}2\zeta
\frac{d n}{d \zeta}+\alpha^2n=\Psi(\zeta),\,\,\,\,\mid \zeta\mid<
1,\,\,\, \alpha^2=\frac{2}{\gamma-1}[(\lambda-m)^2+2-\gamma].
 \label{A1}
\end{equation}
Equation (\ref{A1}) is supplemented by the boundary conditions at
the disc edges. In addition, the solutions of Eq. (\ref{A1})
satisfy the same symmetry as that of the driving term on the right
hand side, namely either symmetric ($\Psi(\zeta)=\Psi(-\zeta)$) or
anti-symmetry ($\Psi(\zeta)=-\Psi(-\zeta)$)  conditions:
\begin{equation}
\frac{dn}{d\zeta}=0\,\,\,\mbox{for}
\,\,\,\zeta=0, \,\,\, n=0\,\,\,\mbox{for}
\,\,\,\zeta= 1,
 \label{A2}
 \end{equation}
\begin{equation}
n=0\,\,\,\mbox{for}
\,\,\,\zeta=0, \,\,\, n=0\,\,\,\mbox{for}
\,\,\,\zeta= 1.
 \label{A3}
\end{equation}
Transforming to a new independent variable  $\theta=\arcsin
\zeta$, problem (\ref{A1})-(\ref{A3}) assumes the following form:
\begin{equation}
\frac{d^2 n}{d \theta^2}-
\frac{3\gamma-5}{\gamma-1}\tan\theta
\frac{d n}{d
\theta}+\alpha^2n=\Psi(\zeta),\,\,\,-\pi/2<\theta<\pi/2,
 \label{A4}
\end{equation}
\begin{equation}
\frac{d n}{d\theta}=0\,\,\,\mbox{for}
\,\,\,\theta=0, \,\,\, n=0\,\,\,\mbox{for}
\,\,\,\theta= \pi/2,
 \label{A5}
 \end{equation}
\begin{equation}
n=0\,\,\,\mbox{for}
\,\,\,\theta=0, \,\,\, n=0\,\,\,\mbox{for}
\,\,\,\theta= \pi/2.
 \label{A6}
\end{equation}
For arbitrary values of $\gamma$  the solutions of Eq. (\ref{A4})
may be expressed in terms of the hypergeometric functions.
However, for the specific value  $\gamma=5/3$  Eq. (\ref{A4})
becomes remarkably simple. Conveniently therefore, considering
from now on the case with   $\gamma=5/3$ as well as Keplerian
rotation, $\Omega=r^{-3/2}$.

\subsection{The eigenvalue problem for the sound mode}
 The eigenvalue problem
for the sound modes is given by (see Eq. (\ref{27}) for
$\gamma=5/3$):
\begin{equation}
\frac{d^2 n}{d \theta^2}-
+\alpha^2n=0,\,\,\,0<\theta<\pi/2,
 \label{A7}
\end{equation}
\begin{equation}
n=0\,\,\,\mbox{for}
\,\,\,\theta= \pm \pi/2.
 \label{A8}
\end{equation}
The general solution of Eq.  (\ref{A7}) is:
\begin{equation}
n=C_1\cos\alpha\theta +C_2\sin\alpha\theta,\,\,
 \label{A9}
\end{equation}
where $\cos\alpha\theta$  and $\sin\alpha\theta$   are the
symmetric and anti-symmetric eigenfunctions. Imposing boundary
conditions (\ref{A5}) and (\ref{A6}), the following two families
of eigenvalues and eigenfunctions obtained:
\begin{equation}
\alpha=2k+1,\,\,\,n=N_{2k+1}\cos(2k+1)\theta,\,\,\,k=0,\pm 1,\pm
2,\ldots,
 \label{A10}
\end{equation}
\begin{equation}
\alpha=2k,\,\,\,\,\,\,\,\,\,\,\,\,\,n=N_{2k}\sin2k\theta,\,\,\,\,\,\,\,\,\,\,\,\,\,\,\,\,\,\,
\,\,\,\,\,\,k=\pm 1,\pm 2,\ldots.
 \label{A11}
\end{equation}

\subsection{The regular symmetric  solution   for the non-resonantly
driven sound mode}
 The following choice of a symmetric driving function
 $\Psi(\zeta)$ in  Eq. (\ref{A4})
is considered (see Eqs. (\ref{38}) and (\ref{40}) with
$\gamma=5/3$):
\begin{equation}
\frac{1}{4}\frac{d^2 n}{d \theta^2}
+n=\frac{1}{4}\Psi(\zeta)\equiv
\psi_3\cos^3\theta+\psi_5\cos^5\theta+\psi_7\cos^7\theta,\,\,\,0<\theta<\pi/2,
\,\,\,
 \label{A12}
\end{equation}
\begin{equation}
\frac{d n}{d\theta}=0\,\,\,\mbox{for}
\,\,\,\theta=0, \,\,\, n=0\,\,\,\mbox{for}
\,\,\,\theta= \pi/2,
 \label{A13}
 \end{equation}
 where  $\lambda=m\pm 1$,  $\alpha^2=4$. The motivation for such a
 choice of the right hand side in eq. (\ref{A12}) is the following:
 recalling solutions
 (\ref{20}) and (\ref{21}) for ${v}_r^{\prime}$
  and ${v}_{\phi}^{\prime}$, as well as definition (\ref{19}), and
  the steady state solution for the number density (\ref{12}) [$\bar{n}\sim (
  1-\zeta^2)^{3/2}\sim
  \cos^3\theta$
  ], it is seen that the
  solution for ${V}_r^{\prime}$
  and ${V}_{\phi}^{\prime}$ that are regular at the disk edges $\zeta ^2 =1$
  are given by a function $G$ that contains powers of third
  order and higher of $\cos\theta$.
  It will consequently be shown in this subsection that similar regularity
  requirements on ${V}_{z}^{\prime}$
  imply that $G$ must contain at least also the next two higher odd powers of
  $\cos\theta$
  whose amplitudes are determined by the arbitrary amplitude of the cubic term.

 The general solution of (\ref{A12})) for symmetric modes is
\begin{equation}
n=
C_2\cos2\theta+C_1\cos\theta+C_3\cos3\theta+C_5\cos5\theta+C_7\cos7\theta,
 \label{A15}
\end{equation}
where from (\ref{A13})  at $\theta =\pi/2$  follows that $C_2=0$,
and
\begin{equation}
C_1=\psi_3+\frac{5}{6}\psi_5+\frac{35}{48}\psi_7,\,\,\,
C_3=-\frac{1}{5}\psi_3-\frac{1}{4}\psi_5-\frac{21}{80}\psi_7,\,\,\,\,
C_5=-\frac{1}{84}\psi_5-\frac{1}{48}\psi_7,\,\,\,
C_7=-\frac{1}{720}\psi_7.\,\,\,
 \label{A16}
\end{equation}
For further convenience (\ref{A15}) is rewritten in the form
\begin{equation}
n= N_1\cos\theta+N_3\cos^3\theta+N_5\cos^5\theta+N_7\cos^7\theta,
 \label{A17}
\end{equation}
where
$$
 N_1=C_1-3C_3+5C_5-7C_7,\,\,\,
 N_3=4C_3-20C_5+56C_7,\,\,\,
 N_5=16C_5-112C_7,\,\,\,
 N_7=64C_7,
 \,\,\,\,\,\,\,\,\,\,\,\,\,\,\,\,\,\,\,\,\,\,\,\,\,\,\,\,\,\,\,\,\,\,\,\,\,\,\,\,\,\,
 \,\,\,\,\,\,\,\,\,\,\,\,\,\,\,\,\,\,\,\,\,\,\,\,\,\,\,\,\,\,\,\,\,\,\,\,\,\,\,\,\,\,
 \,\,\,\,\,\,\,\,\,\,\,\,\,\,\,\,\,\,\,\,\,\,\,\,\,\,\,\,\,\,\,\,\,\,\,\,\,\,\,\,\,\,
 \,\,\,\,\,\,\,\,\,\,\,\,\,\,\,\,\,\,\,\,\,\,\,\,\,\,\,\,\,\,\,\,\,\,\,\,\,\,\,\,\,\,
 $$
or using (A16)
\begin{equation}
N_1=\frac{8}{5}\psi_3+\frac{32}{21}\psi_5+\frac{64}{45}\psi_7,\,\,\,
N_3=-\frac{4}{5}\psi_3-\frac{16}{21}\psi_5-\frac{32}{45}\psi_7,\,\,\,\,
N_5=-\frac{4}{21}\psi_5-\frac{8}{45}\psi_7,\,\,\,
N_7=-\frac{4}{45}\psi_7.\,\,\,
 \label{A18}
\end{equation}
Following  the regularity requirement on ${V}_{z}^{\prime}$ and
Eqs. (\ref{39})  the coefficient of
$\cos\theta\sim(1-\zeta^2)^{1/2}$ in  (\ref{A17}) must is zero,
namely:
\begin{equation}
N_1=\frac{8}{5}\psi_3+\frac{32}{21}\psi_5+\frac{64}{45}\psi_7=0.\,\,\,
 \label{A19}
\end{equation}

Applying relations (\ref{A12}) -(\ref{A19})  to Eq. (\ref{38})
 for $\hat{n}^{(1)}$ yields:
\begin{equation}
G= G_3\cos^3\theta+G_5\cos^5\theta+G_7\cos^7\theta,
 \label{A20}
\end{equation}
and
\begin{equation}
\psi_j^{(1)}=\mp i\frac{9}{8}\lambda G_j,\,\,\,\,\lambda=m\pm 1.
 \label{A21}
\end{equation}
Here $\psi_j^{(1)}$  are the coefficient for  the right hand side
of Eq. (\ref{38}) for $\hat{n}^{(1)}$, the coefficients  $G_j$
will be specified below ($j=3,5,7$). The above relations determine
$\hat{n}^{(1)}$
\begin{equation}
\hat{n}^{(1)}=
N_1^{(1)}\cos\theta+N_3^{(1)}\cos^3\theta+N_5^{(1)}\cos^5\theta+N_7^{(1)}\cos^7\theta,
 \label{A22}
\end{equation}
where $N_j^{(1)}$  are expressed in terms of $\psi_j^{(1)}$,
($j=1,3,5,7$), through relations (\ref{A18})
\begin{equation}
N_1^{(1)}=\mp i \lambda
(\frac{9}{5}G_3+\frac{12}{7}G_5+\frac{8}{5}G_7),\,\,\,\,
N_3^{(1)}=\pm i \lambda
(\frac{9}{10}G_3+\frac{6}{7}G_5+\frac{4}{5}G_7),\,\,\,\,
N_5^{(1)}=\pm i \lambda(\frac{3}{14}G_5+\frac{1}{5}G_7),\,\,\,
N_7^{(1)}=\pm i \lambda\frac{1}{10}G_7,\,\,\,
 \label{A23}
\end{equation}
with the regularity condition
\begin{equation}
N_1^{(1)}=\mp i \lambda
(\frac{9}{5}G_3+\frac{12}{7}G_5+\frac{8}{5}G_7)=0,\,\,\,\,
\lambda=m\pm 1.
 \label{A24}
\end{equation}
Hence  $\hat{n}^{(1)}$ is determined through $G_j$  ($j=3,5,7$),
which will be specified below.

Applying relations (\ref{A12}) -(\ref{A19})  to Eq. (\ref{40})
 for $\hat{n}^{(0)}$ yields:
\begin{equation}
\psi_j^{(0)}=\mp\frac{3}{2}N_j^{(1)}-\frac{3}{4}i(m\pm 2)
G_j,\,\,\,\,(j=1,3,5,7).
 \label{A25}
\end{equation}
Substituting (\ref{A23}) into (\ref{A25}) yields
$$
\psi_1^{(0)}=-\frac{3}{4}i(m\pm 2) G_1,\,\,\,\,
\psi_3^{(0)}=-\frac{3}{20}i(14 m \pm 19) G_3-\frac{9}{7}i(m \pm 1)
G_5-\frac{6}{5}i(m \pm 1) G_7,
\,\,\,\,\,\,\,\,\,\,\,\,\,\,\,\,\,\,\,\,\,\,\,\,\,\,\,\,\,\,\,\,\,\,\,\,\,\,\,\,\,\,\,\,\,\,\,\,\,\,\,\,
\,\,\,\,\,\,\,\,\,\,\,\,\,\,\,\,\,\,\,\,\,\,\,\,\,\,\,\,\,\,\,\,\,\,\,\,\,\,\,\,\,\,\,\,\,\,\,\,\,\,\,\,
\,\,\,\,\,\,\,\,\,\,\,\,\,\,\,\,\,\,\,\,\,\,\,\,\,\,\,\,\,\,\,\,\,\,\,\,\,\,\,\,\,\,\,\,\,\,\,\,\,
\,\,\,\,\,\,\,\,\,\,\,\,\,\,\,\,\,\,\,\,\,\,\,\,\,\,\,\,\,\,\,\,\,\,\,\,\,\,\,\,\,\,\,\,\,\,\,\,\,\,\,\,\,\,\,
$$
\begin{equation}
 \psi_5^{(0)}=-\frac{3}{28}i(10 m \pm 17) G_5-\frac{3}{10}i(m \pm
1) G_7,\,\,\,\,
\psi_7^{(0)}=-\frac{3}{20}i(6 m \pm 11) G_7,
 \label{A26}
\end{equation}
and
\begin{equation}
\hat{n}^{(0)}=
N_1^{(0)}\cos\theta+N_3^{(0)}\cos^3\theta+N_5^{(0)}\cos^5\theta+N_7^{(0)}\cos^7\theta.
 \label{A27}
\end{equation}
The coefficients $N_j^{(0)}$  are expressed through $\psi_j^{(0)}$
according to (\ref{A18}) and then by using (\ref{A26}) through
$G_j$, ($j=1,3,5,7$). Once again, requiring the regularity of
$V^{\prime}_z$ and employing Eq. (\ref{41}) results in:
\begin{equation}
N_1^{(0)}=- i \frac{6}{25}(14m\pm 19)G_3-i\frac{8}{245}(113m\pm
148)G_5-i\frac{16}{175}(40m\pm 37)G_7=0.
 \label{A28}
\end{equation}
Finally, setting, e.g. $G_3=1$ as an arbitrary amplitude of the
inertia-coriolis driving waves, Eqs. (\ref{A24}) and (\ref{A28})
determine the coefficients $G_5$ and $G_7$. Since both $F(r)$ and
$G(\zeta)$ are known, the corresponding velocities $\hat{v}_r$,
$\hat{v}_\phi$, $\hat{v}_z^{(k)}$ ($k=0,1$) may be calculated
along with the number density.

\bsp
\label{lastpage}
\end{document}